\documentclass[aps,prl,twocolumn,superscriptaddress,preprintnumbers,showpacs, amsmath,amssymb]{revtex4-1}

\usepackage{graphicx}
\usepackage{dcolumn}
\usepackage{bm}
\usepackage{hyperref}
\usepackage[usenames,dvipsnames]{color}
\usepackage[normalem]{ulem}
\usepackage{amsmath}
\usepackage{amsfonts}
\usepackage{multirow}
\usepackage{times}

\newcommand{\comment}[1]{\textcolor{red}{#1}}
\renewcommand{\comment}[1]{\relax}

\newcommand{\todelete}[1]{\textcolor{green}{\sout{#1}}}
\renewcommand{\todelete}[1]{\relax}

\usepackage{ulem}
\usepackage[mathlines]{lineno}

\begin{document}

\title{Electronic properties of ultra-thin YCrO$_3$  films}

\author{Banabir Pal}
\email{bp435@physics.rutgers.edu}
\affiliation{Department of Physics and Astronomy, Rutgers University, Piscataway, New Jersey 08854, USA}
\author{Xiaoran Liu}
\email{xiaoran.liu@rutgers.edu}
\affiliation{Department of Physics and Astronomy, Rutgers University, Piscataway, New Jersey 08854, USA}
\author{Fangdi Wen}
\affiliation{Department of Physics and Astronomy, Rutgers University, Piscataway, New Jersey 08854, USA}
\author{Mikhail Kareev}  
\affiliation{Department of Physics and Astronomy, Rutgers University, Piscataway, New Jersey 08854, USA}
\author{A. T. N'Diaye}
\affiliation{Advanced Light Source, Lawrence Berkley National Laboratory, Berkeley, California 94720, USA}
\author{P. Shafer}
\affiliation{Advanced Light Source, Lawrence Berkley National Laboratory, Berkeley, California 94720, USA}
\author{E. Arenholz}
\affiliation{Advanced Light Source, Lawrence Berkley National Laboratory, Berkeley, California 94720, USA}
\author{Jak Chakhalian}
\affiliation{Department of Physics and Astronomy, Rutgers University, Piscataway, New Jersey 08854, USA}
\date{\today}

\begin{abstract}
We report on the heteroepitaxial stabilization of  YCrO$_3$ ultra-thin films on LSAT (001) substrate. Using a combination of resonant X-ray absorption spectroscopy (XAS) and atomic multiplet cluster calculation, the electronic structure of YCrO$_3$ thin film was investigated. Polarization dependent Cr L$_{3,2}$ edge XAS measurement reveal the presence of  an anomalous orbital polarization uncharacteristic of a 3d$^3$ electronic system. Atomic multiplet calculation demonstrate the critical importance of charge transfer energy, coulomb correlation strength and hopping interaction in stabilizing this  unusual orbital polarized state likely connected to  the bulk  multiferroicity. 
\end{abstract}

\maketitle
\indent Complex oxides with non-collinear magnetic structures have emerged as a forefront area in condensed matter physics for envisioning and investigating multiferroic behavior associated with large magnetoelectric couplings \cite{ref19,Eric_JPCM_2016,B1, B2, B3}. Recently, YCrO$_3$ (YCO) has been discovered as a new  member of this category, which undergoes a ferroelectric transition at around 473 K, followed by a canted antiferromagnetic transition at 140 K resulting in weak ferromagnetism \cite{ref18}. The presence of the multiferroic feature in YCO rapidly promoted it as a potential candidate for future multifunctional applications \cite{Sahu_JMC_2007, Cheng_JAP_2010}. Towards this goal, synthesis of high-quality epitaxial YCO thin films is a key  step\cite{XL1}.\\           
\indent Despite several past attempts to obtain structurally and stoichiometrically well-defined YCO film it turns out to be rather challenging work. Specifically, since Cr$^{3+}$ can be further oxidized to higher valence states, oxygen-rich secondary phases can emerge \cite{Long_PRB_2007}. Moreover, the choice of substrate is important to eliminate the formation of polycrystalline  films \cite{Seo_JCC_2013, Araujo_SciRep_2014}. To date, layer-by-layer growth  of high-quality epitaxial YCO\  films has not been demonstrated. Considering that YCO belongs to the large family of perovskite rare-earth chromites RCrO$_3$ (R = Ho, Er, Yb, Lu and Y), all of which in the bulk exhibit the similar multiferroic behavior \cite{Eric_JPCM_2016, Sahu_JMC_2007}, a systematic investigation of the growth of YCO ultra-thin film will build a link for the fabrication of those compounds in thin film form.\\     
\indent In this letter,  we show that by judicious control of oxygen pressure and using a suitable post-annealing process ultra-thin epitaxial YCO films can  be stabilized. X-ray diffraction (XRD), reflectivity (XRR), and photo-electron spectroscopy (XPS) confirm their excellent  crystallinity with proper stoichiometry and chemical valence. The electronic structure of YCO was further investigated by means of resonant X-ray absorption spectroscopy on Cr L-edge and elucidated  by multiplet cluster calculations. Polarization dependent XAS measurements on Cr L$_{3,2}$ edge reveal the presence of an unexpectedly strong dichroic  signal signifying a large degree of orbital polarization highly atypical of 3d$^3$ systems. \\
\indent The schematic crystal structure of YCO along the $b$ axis and with in $ac$ plane is shown in Fig. 1(a) and 1(b), respectively. 
\begin{figure}[bt]
\begin{center}
\includegraphics[width=1.0\columnwidth]{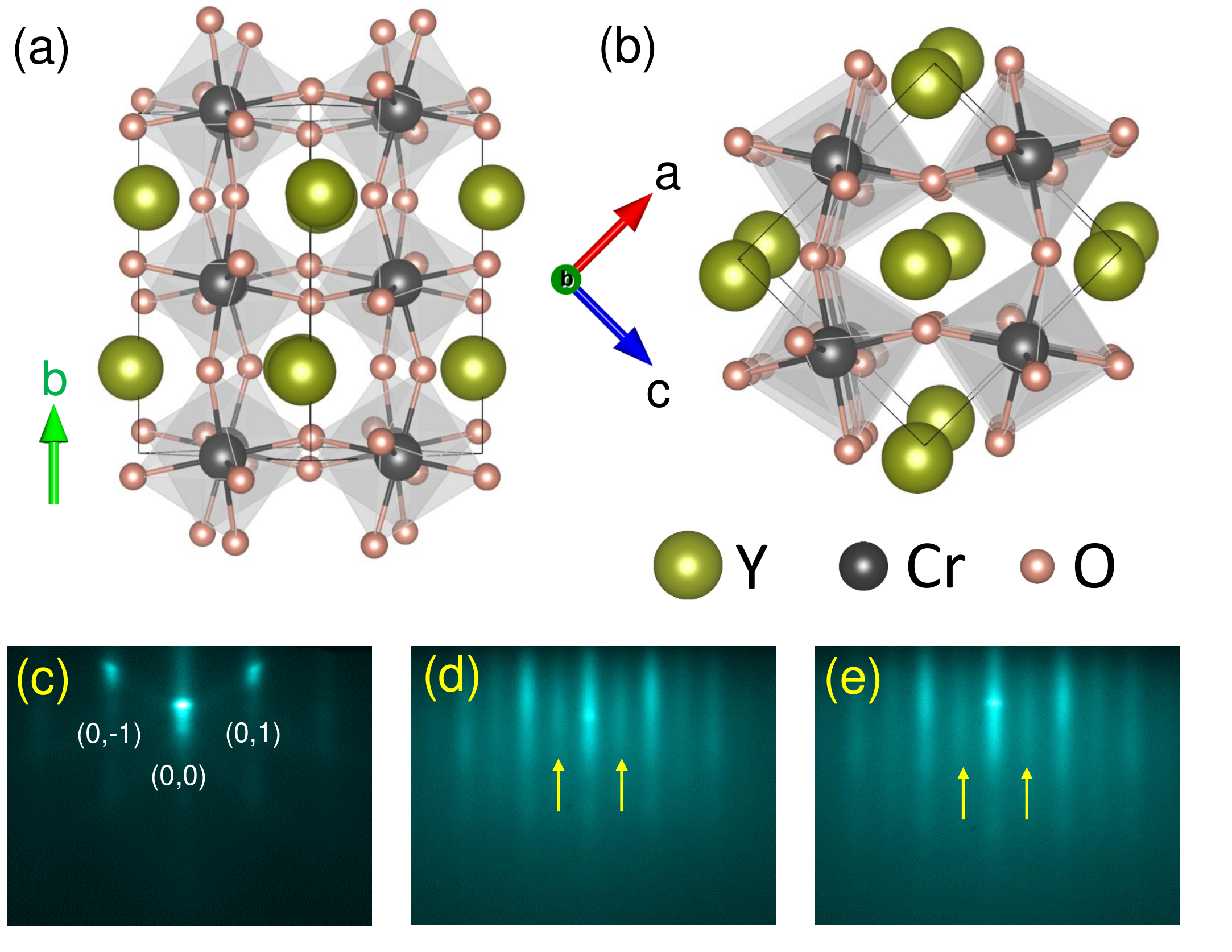}
\caption{Schematic representation of YCrO$_3$ crystal structure with space group $Pnma$. Fig. 1(a) and 1(b) shows the arrangement of atoms along the $b$ axis and within the $ac$ plane, respectively. RHEED images of pure (001) LAST substrate, YCrO$_3$ films after depositions, and YCrO$_3$ after the annealing are shown in Fig. 1(c), (d) and (e) respectively. Half order peaks in Fig. 1(d) and 1(e) is highlighted with yellow arrows.}
\label{fig1}
\end{center}
\end{figure}  
Since the size of Y$^{3+}$ ion is smaller as compared to the void created by CrO$_6$ octahedra,  YCO naturally crystallizes in to a distorted orthorhombic perovskite  structure with space group $Pnma$ and lattice parameters $a$ = 5.5226 \AA, $b$ = 7.53555 \AA, $c$ = 5.24376 \AA. \cite{ref22,ref23} Starting from the ideal cubic perovskite symmetry, the crystal structure of YCO is obtained by an in-phase tilting of the adjacent CrO$_6$ octahedra along the pseudo-cubic $b$ axis, and anti-phase tilting along both $a$ and $c$ axes \cite{ref24} (described as $a^{-}b^{+}a^{-}$ from Glazer notation\cite{ref25}).\\  
\indent Epitaxial YCO thin films  were grown layer-by-layer on LSAT (001) substrates by laser MBE using a KrF excimer laser operating at $\lambda$ = 248 nm (repetition rate of 2 Hz with a laser fluence close to 2 J$\cdot$cm$^{-2}$). The substrate temperature was maintained at 750$^{\circ}\mathrm{C}$ and the deposition process was monitored by {\it{in-situ}} reflection-high-energy-electron-diffraction (RHEED). Representative RHEED patterns of LSAT (001) surface, YCO film after the deposition, and YCO film after the annealing process are shown in Fig. 1.(c), 1(d), and 1(e), respectively. Both the specular and the off-specular reflections on the first Laue zone are clearly observed throughout the entire growth, indicative of the flatness of YCO films with good crystallinity. 
\begin{figure}[t]
\begin{center}
\includegraphics[width=0.85\columnwidth]{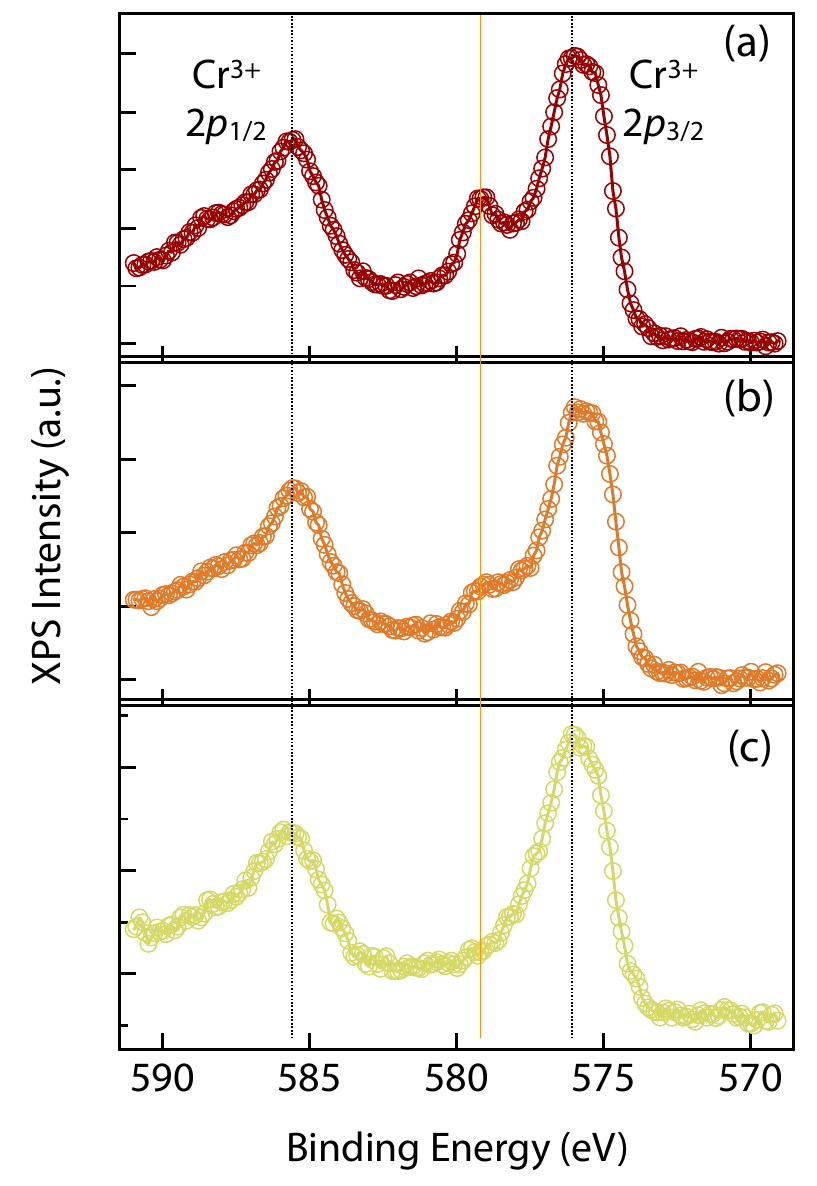}
\caption{(a)-(c) Cr 2$p$ core level XPS spectra of YCO films grown under different conditions. The peaks associated with Cr$^{3+}$ (black dashed lines) and its higher oxidation state (orange line) are labeled on the figure, respectively.}
\label{fig2}
\end{center}
\end{figure}
Additionally, the presence of distinct half-order reflections  highlighted by the yellow arrows in Fig. 1(d) and 1(e) confirm that the expected orthorhombic distortions persist in our  YCO thin films\cite{ref26}.\\
\indent The charge states and the stoichiometry of the YCO thin films were investigated using core-level XPS measurements in a Thermo Scientific X-ray Photoelectron Spectrometer equipped with a monochromatic Al K$_\alpha$ (1486.6 eV) source. The photoelectrons were collected with a hemispherical analyzer (energy resolution $\sim$450 meV) in a surface normal geometry  to increase the bulk sensitivity of the measurements. The energy of each spectrum was calibrated using C 1$s$ reference spectrum to compensate for a possible charging effect.\\ 
\indent Figure 2 shows Cr 2$p$ XPS spectra of several YCO thin films fabricated under different oxygen partial pressure. For YCO grown at a relatively high oxygen pressure of $\sim 10^{-1}$ Torr, the Cr 2$p$ spectrum shown in Fig. 2(a) exhibits two main peaks at binding energies of $\sim$ 576 eV and $\sim$ 585.6 eV, characteristic of the 2p$_{3/2}$ and 2p$_{1/2}$ doublet feature of Cr$^{3+}$ with a spin-orbit splitting strength close to 9.6 eV. \cite{ref27new} Moreover, the spectrum also contains two additional peaks around $\sim$ 579.2 eV and $\sim$ 588.6 eV, as indicated by the orange lines on the figure. Comparing the chemical shift associated with these extra features to previously reported  results \cite{ref27, ref28, ref29}, we assign  these peaks to the higher Cr$^{6+}$ oxidation state. These results imply that in addition to the proper phase of YCO, another oxygen-rich secondary phases (YCrO$_{3+\delta}$) has been stabilized in the film grown at $\sim 10^{-1}$ Torr.\\ 
\indent The formation of the secondary phase can be effectively suppressed by growth at a lower oxygen pressure. 
\begin{figure}[b]
\begin{center}
\includegraphics[width=1.1\columnwidth]{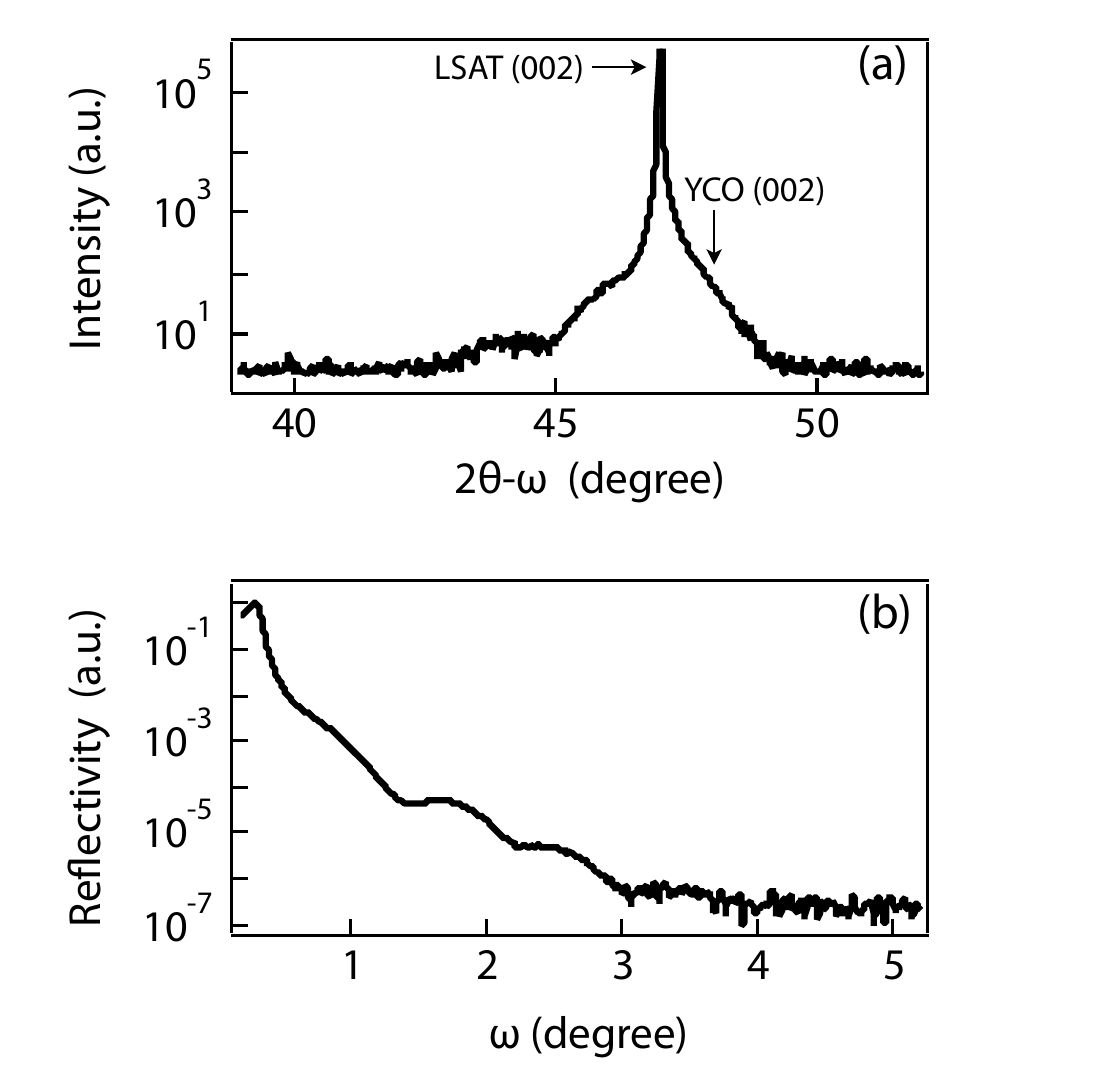}
\caption{(a) X-ray diffraction result of YCO thin film. The sharp peak around 2$\theta$ $\sim$ 47$^{\circ}$ represents the (002) reflection of the LSAT substrate. (b) X-ray reflectivity curves of YCO thin film.}
\label{fig3}
\end{center}
\end{figure}
A new series of YCO samples were synthesized at  oxygen partial pressure  of $\sim 10^{-3}$ Torr. The corresponding Cr 2$p$ spectrum shown  in Fig. 2(b) exhibits a significant drop in Cr$^{6+}$ intensity. However, further lowering the oxygen pressure down to $\sim 10^{-5}$
Torr significantly
affects the growth kinetics and results in three dimensional (3D) growth.  Based on this observation to completely remove the secondary phases, we grew the YCO films at $\sim 10^{-3}$ Torr oxygen pressure and post annealed in vacuum for 15 min, followed by cooling in vacuum down to  room temperature.  XPS data corresponding to  this growth process are shown in Fig.2(c). As immediately seen, the spectrum is composed of a doublet feature from the pure Cr$^{3+}$ charge state. Additionally, cross section corrected intensity ratio of the Cr 2$p$ and Y 3$d$ were calculated to estimate the Y/Cr ratio in this optimized system. The result is found to be 1.1$\pm$0.2, further corroborating the expected chemical composition of the film.\\
\indent In order to ensure the single phase synthesis, structural properties of each YCO thin film were characterized by high-resolution XRD measurements (Cu $K_{\alpha1}$  radiation source, $\lambda$ = 1.5406 \AA). Figure 3(a) shows the high-resolution XRD 2$\theta$-$\omega$ scans in the vicinity of the (002) reflection of LSAT substrate. No impurity peak is observed signifying the formation of single phase YCO. In addition, thickness of the film was determined from the XRR measurements shown in Fig.3(b). Based on the interval of neighboring thickness fringes, the total thickness of the YCO film is determined to be close to 5.7 nm, which is in excellent agreement  with  the thickness predicted from RHEED oscillations.\\ 
\indent After establishing the successful growth mode for high quality YCO ultra-thin films, their electronic structure was investigated by  polarization dependent XAS measurements at beamline 4.0.2 of the Advanced Light Source in Lawrence Berkeley National Laboratory. Each XAS spectrum was acquired in the luminescence detection mode at 15 K. The angle of X-ray incidence to the sample surface is 15$^{\circ}$. The relative orientation of the polarization vector (both in-plane and out-of-plane) of the incoming photon with respect to the film is  schematically shown in  inset of Fig. 4(a). Polarization dependent XA can probe local orbital character  known as linear dichroism; as such this probe is very sensitive to the sub-band splitting of the degenerate $t_{2g}$ states and orbital anisotropy. Experimentally obtained polarization dependent Cr L$_{2,3}$ XAS spectra for YCrO$_3$ (both in-plane and out-of-plane) are displayed in Fig. 4(a). The multiplet features in energy range from 573 eV to 583 eV are part of the L$_3$ edge whereas those from 583 eV to 593 eV belong to the L$_2$ edge. 
\begin{figure}[h!t]
\begin{center}
\includegraphics[width=1\columnwidth]{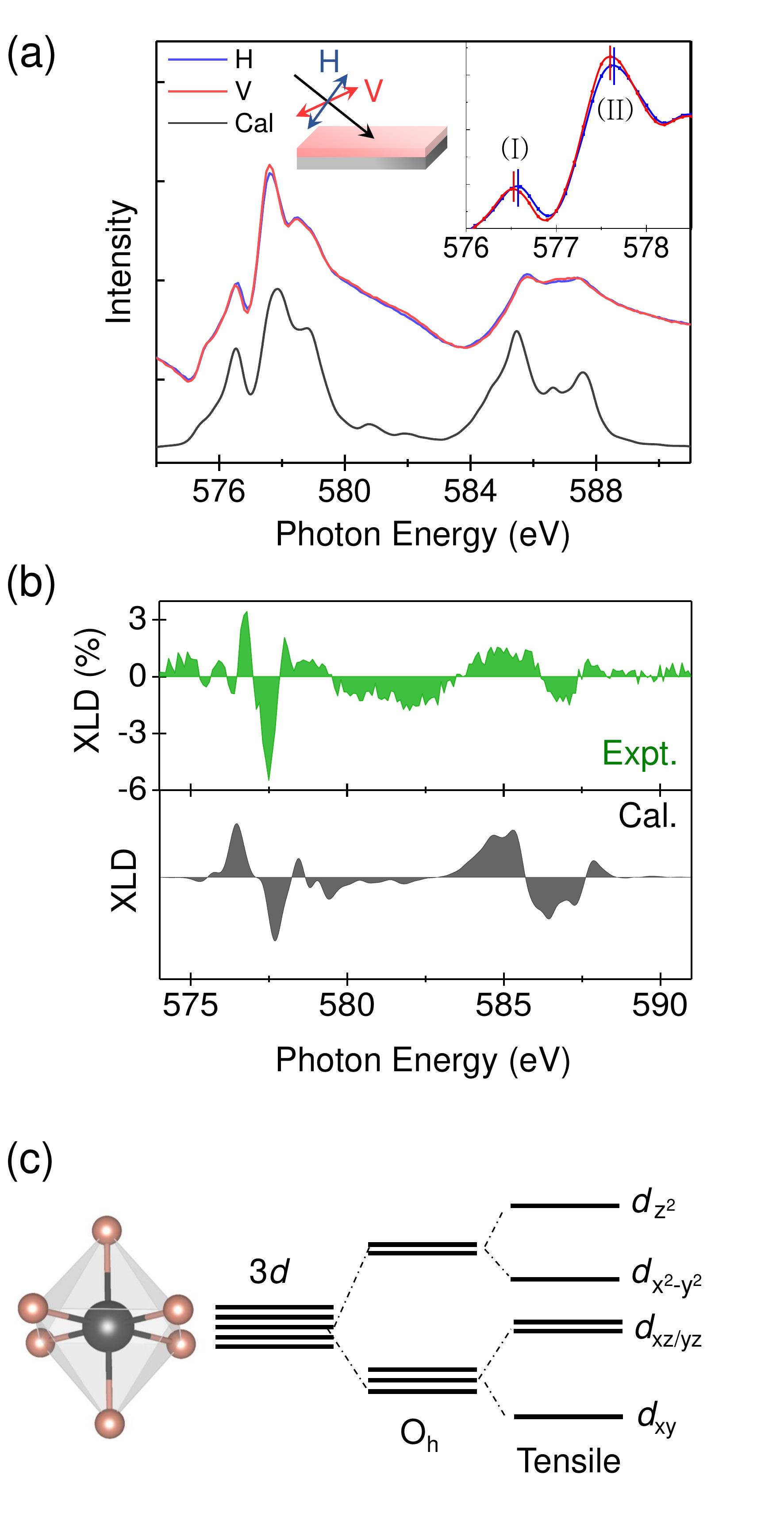}
\caption{(a) Both in-plane (V) and out-of-plane (H) polarization dependent XAS spectra is shown along with calculated XAS spectrum. Schematic view of the polarization vector of the incoming photon is shown in the first inset. In-plane and out-of-plane polarization directions in this figure corresponds to vertical and horizontal polarization, respectively as mentioned in the main text. Second inset represent the enlarge version of the XAS measurement near L$_3$ edge. (b) The difference between in-plane and out-of-plane spectra or XLD signal is shown as the green shaded area in the top panel. Bottom panel contains calculated XLD signal as black shaded area.  (c) The schematic representation of the splitting of 3$d$ states under tensile octahedral crystal field.}
\label{fig4}
\end{center}
\end{figure}
To interpret the origin of the multiple peak structure in L$_{3,2}$ edges, atomic multiplet cluster calculations were carried out using the charge transfer multiplet method (CTM4XAS) \cite{ref31,ref31A}. A CrO$_6$ cluster with Cr$^{3+}$ charge state and octahedral crystal symmetry ($O_h$) were used to simulate the XAS spectra  presented in Fig. 4 (black curve). The optimized atomic parameters for the description of the experimental data are summarized in Table I. 
The remarkable reproduction of all main features of the simulated spectrum clearly suggests that Cr is indeed in the Cr$^{3+}$ oxidation state which is in agreement with the XPS spectrum. Moreover, large positive charge transfer energy ($\Delta$=2.5 eV) signifies that the ground state is dominated by the $d^3$ electronic state. The $d^4$\underline{L} and $d^5$\underline{L}$^2$ states can be excluded because the difference in energy between $d^3$ and $d^4$\underline{L}/$d^5$\underline{L}$^2$ states are expressed as E($d^4$\underline{L})-E($d^3$)=$\Delta$ and E($d^5$\underline{L}$^2$)-E($d^3$)=$\Delta$+$U_{dd}$, both  are positive in our calculations. Moreover, the higher magnitude of  charge transfer energy ($\Delta$=2.5 eV)  compared to the Coulomb correlation strength (U$_{dd}$=1.0 eV) suggests that this system is in  the Mott-Hubbard insulating regime.\\
\setlength{\tabcolsep}{0.6em}
{\renewcommand{\arraystretch}{1.3}
\begin{table}
\centering
\caption{Best atomic parameters obtained from multiplet cluster calculations of Cr$^{3+}$ in O$_h$ symmetry. All energy parameters are given in eV.}
\label{my-label}
\begin{tabular}{c c c c c c}
\hline\hline
10Dq  & $U_{dd}$-$U_{pd}$  & $F_{dd}, F_{pd}, G_{pd}$ & $\Delta$   & $E_g$ & $T_{2g}$ \\[1ex] 
\hline
1.8   & 1.00             & 1.0, 1.0, 1.0      & 2.5   &  2.0   &  1.1     \\ 
\hline\hline
\end{tabular}
\end{table}
\indent Normally, one would expect no orbital polarization in this system since the formal oxidation state of Cr in YCO is Cr$^{3+}$ (3$d^3$). Therefore, YCO is Jahn-Teller inactive leading to almost no dispersion in the Cr-O bond length as compared to JT-active manganites. Surprisingly, a large XLD signal of $\sim 6\%$ was observed in our experiments as shown in Fig. 4(b). In addition,  a small shift in the peak position in the two main features (feature I and II) of the L$_3$ edge was found in the XLD\ spectra and is  shown in  the inset of Fig. 4(a). To understand this surprisingly result, atomic multiplet cluster calculations were carried out using same CTM4XAS software. Interestingly, the calculation only within crystal field approximation excluding charge transfer energy ($\Delta$), on-site Coulomb interaction strength ($U_{dd}$) and hoping interaction (T$_{T_{2g}}$ and T$_{e_g}$) does not exhibit any XLD signal. This clearly implies that crystal field alone cannot capture the electronic structure of YCO. On the other hand a set of parameters shown in Table I can successfully reproduce the XLD spectra thus stressing  the importance of covalency for the electronic structure.\\
\indent Although the multiplet cluster calculations could reproduce all the spectral features, it is possible that some other effects can be responsible for this large XLD signal. One such effect can arise from the tensile strain (+1.5\%) present in our system.  Tensile strain can break the local octahedral symmetry to create an intriguing orbital polarization even in a 3d$^3$ electronic system. The schematic view of the energy splitting due to tensile strain is presented in Fig. 4(c). In the case of tensile strain, the four oxygen ions located in  the $xy$ plane move away from the the central atom, resulting in the stabilization of the $d_{xy}$ and $d_{x^2-y^2}$ states as compared to $d_{xz/yz}$ and $d_{z^2}$ states (i.e. $d_{xy}$ $<$ $d_{xz/yz}$ $<$ $d_{x^2-y^2}$ $<$ $d_{z^2}$ ) \cite{ref32}. The observed energy splitting of the feature I and II at the L$_3$ edge  (see inset of the Fig. 4(a)) may further support this argument. The cumulative effect of covalency, metal-ligand hybridization and tensile strain can be responsible for this anomalously large XLD signal in the YCO ultra thin films.\\  
\indent In summary, high quality YCO thin films were epitaxially grown on LSAT (001) substrates in a layer-by-layer growth mode. The oxygen partial pressure during deposition and the post-annealing process are found to play a crucial role in stabilizing single phase YCO films with proper Cr$^{3+}$ charge state and stoichiometry. The quality of the sample fabricated under the optimized condition has been verified by XPS, XRD and XRR characterizations. Polarization dependent XAS measurements demonstrate strong \textit{d}-electron orbital polarization. Multiplet cluster calculations which include  optimized charge transfer energy, electron-electron correlations and hopping interaction  successfully  reproduced the XLD spectra.  In addition we conjecture that tensile strain might also play an important  role in stabilizing the large orbital polarization that may shed light on the origin of  multiferroic state in YCO.  \\

\textbf{Acknowledgement:} The authors  deeply thank J. Rondinelli and D. Puggioni for fruitful discussions. J.C., X.L. and B.P. are supported by the Gordon and Betty Moore Foundation EPiQS Initiative through Grant No. GBMF4534 for  synthesis effort and XPS\ characterization work at  Rutgers. F.W. is supported by the U.S. Department of Energy (DOE) under Grant No. DOE DE-SC 00012375 for  characterization work  at  synchrotron.  This research used resources of the Advanced Light Source, which is a Department of Energy Office of Science User Facility under Contract No. DE- AC0205CH11231.


\begin{thebibliography}{999}
\bibitem{ref19}
R. Ramesh and N. A. Spaldin, Nature Materials \textbf{6}, 21–29 (2007)

\bibitem{Eric_JPCM_2016}
E. Bousquet, and A. Cano, J. Phys.: Condens. Matter \textbf{28}, 123001 (2016).

\bibitem{B1}
W. Eerenstein, N. D. Mathur and J. F. Scott, Nature \textbf{442}, 759 (2006)

\bibitem{B2}
N. Hur, S. Park, P. A. Sharma, J. S. Ahn, S. Guha and S-W. Cheong, Nature \textbf{429}, 392 (2004)


\bibitem{B3}
T. Zhao, A. Scholl, F. Zavaliche, K. Lee, M. Barry, A. Doran, M. P. Cruz, Y. H. Chu, C. Ederer, N. A. Spaldin, R. R. Das, D. M. Kim, S. H. Baek, C. B. Eom and R. Ramesh, Nature Materials \textbf{5}, 823 (2006)



\bibitem{ref18}
C. R. Serrao, A. K. Kundu, S. B. Krupanidhi, U. V. Waghmare, and C. N. R. Rao, Phys. Rev. B \textbf{72}, 220101(R) (2005)

\bibitem{Sahu_JMC_2007}
J. R. Sahu, C. R. Serrao, N. Ray, U. V. Waghmare, and C. N. R. Rao, J. Mater. Chem. \textbf{17}, 42 (2007).

\bibitem{Cheng_JAP_2010}
Z. X. Cheng, X. L. Wang, S. X. Dou, H. Kimura, and K. Ozawa, J. Appl. Phys. \textbf{107}, 09D905 (2010).

\bibitem{XL1}
X. Liu, S. Middey, Y. Cao, M. Kareev, and J. Chakhalian, MRS Commun. \textbf{6}, 133  (2016).




\bibitem{Long_PRB_2007}
Y. W. Long, L. X. Yang, Y. Yu, F. Y. Li, R. C. Yu, and C. Q. Jin, Phys. Rev. B \textbf{75}, 104402 (2007).

\bibitem{Seo_JCC_2013}
J.-D. Seo, and J. Y. Son, J. Cryst. Growth \textbf{375}, 53 (2013).

\bibitem{Araujo_SciRep_2014}
C. Moyses Araujo \textit{et al.}, Sci. Rep. \textbf{4}, 4686 (2014).


\bibitem{ref22}
J. Prado-Gonjal, R. Schmidt, J.-J. Romero, D. Ávila, U. Amador, and E. Morán, Inorg. Chem., \textbf{52}, 313–320 (2013)


\bibitem{ref23}
S. Kovachev, D. Kovacheva, S. Aleksovska, E. Svab, and K. Krezhov, AIP Conference Proceedings \textbf{1203}, 199 (2010)


\bibitem{ref24}
M. Weber, J. Kreisel, P. A. Thomas, M. Newton, K. Sardar, R. I. Walton, Phys. Rev. B \textbf{85}, 054303 (2012)


\bibitem{ref25}
A. M. Glazer, Acta Cryst. B \textbf{28}, 3384 (1972). 

\bibitem{ref26}
A. Ichimiya and P. I. Cohen, Reflection High-Energy Electron Diffraction (Cambridge University Press, Cambridge, 2004)



\bibitem{ref27new}
W. Yao, T. Duan, Y. Li, L. Yang and K.Xie, New J. Chem., \textbf{39}, 2956 (2015)

\bibitem{ref27}
Y. Chen, K. Ding, L. Yang, B. Xie, F. Song, J. Wan, G. Wang, and M. Han, Appl. Phys. Lett. \textbf{92}, 173112 (2008);


\bibitem{ref28}
M. C. Biesinger, C. Brown, J. R. Mycroft, R. D. Davidson and N. S. McIntyre, Surf. Interface Anal. \textbf{36} 1550 (2004)


\bibitem{ref29}
B. A. Manning, J. R. Kiser, H. Kwon, and S. R. Kanel, Environ. Sci. Technol. \textbf{41}, 586 (2007) 



\bibitem{ref31}
E. Stavitski and F. M. F. de Groot, Micron \textbf{41}, 68 (2010)

\bibitem{ref31A}
H. Ikeno, F. M. F. de Groot, E. Stavitski and I.Tanaka, J. Phys.: Condens. Matter \textbf{21}, 104208 (2009).


\bibitem{ref32}
Y. Cao, X. Liu, P. Shafer, S. Middey, D. Meyers, M. Kareev, Z. Zhong, J. W. Kim, P. J. Ryan, E. Arenholz and J. Chakhalian, npj Quantum Materials \textbf{1}, 16009 (2016)

\end{thebibliography}
\end{document}